\documentclass[conference]{IEEEtran}
\usepackage[utf8]{inputenc}    
\usepackage[T1]{fontenc}       
\usepackage{textcomp}          

\usepackage{graphicx}
\usepackage{algorithm2e}

\usepackage{subcaption}

\begin{document}
%
\title{Community Detection in Multilayer Networks: Challenges, Opportunities and Applications}

\author{\IEEEauthorblockN{Randa Boukabene, Fatima Benbouzid-Si Tayeb}
\IEEEauthorblockA{Laboratoire des Methodes de Conception de Systèmes (LMCS)\\
Ecole nationale Supérieure d’Informatique (ESI)\\
BP 68M - 16270 Oued Smar, Alger, Algeria\\
{\{fr\_boukabene, f\_sitayeb\}@esi.dz}
}}


%


\maketitle

\begin{abstract}
Community detection is a fascinating and rapidly evolving field, but when it comes to analyzing networks with multiple types of interactions—referred to as multilayer networks—there’s still a lot of untapped potential. Despite the wide array of methods developed to identify community structures in such networks, this area remains underexplored, leaving plenty of room for innovation. A systematic review of recent advancements is essential to understand where the field stands and where it’s headed. While significant strides have been made across various disciplines, many questions remain unanswered, and new opportunities are waiting to be uncovered. In this paper, we explore the different types of multilayer networks, community detection techniques, and how they are applied in real-world scenarios. We also dive into the key challenges researchers face and suggest potential directions for future work, aiming to refine community detection techniques and boost their effectiveness in multilayer networks.
\end{abstract}



%
\IEEEpeerreviewmaketitle

\section{\uppercase{Introduction}}
\label{sec:introduction}
Networks have emerged as a powerful tool for modeling the complex relationships within systems across a wide range of fields, from sociology and biology to computer science \cite{boukabene2023improved}. By representing entities as nodes and their interactions as edges, network analysis offers valuable insights into the structural and functional dynamics of these systems. A central focus of this analysis is community detection—the process of identifying groups of nodes that are more densely connected to each other than to the rest of the network. Traditionally, this task has been applied to single-layer networks, which consider only one type of relationship at a time. However, real-world systems are rarely so simple. They often involve multiple, distinct types of interactions. For example, social networks can include layers of family ties, professional relationships, and friendships, each representing a different dimension of human interaction \cite{Boukabene2025}. To better capture this complexity, researchers have turned to multilayer networks \cite{boccaletti2014structure}, where each layer represents a specific type of relationship. This approach provides a more nuanced, multidimensional view of network structure \cite{kivela2014multilayer}.

Detecting communities in multilayer networks is far more challenging than in single-layer networks. Each layer may have its own unique community structure, and these structures don’t always align across layers. This misalignment makes it difficult to identify a unified community structure that accurately reflects the network as a whole. To tackle this problem, researchers have developed a variety of methods aimed at finding a composite community partition that balances the diverse patterns observed across layers \cite{pizzuti2017evolutionary}. These efforts are crucial for advancing our understanding of the intricate, layered interactions that define real-world systems.

Several recent surveys have organized and reviewed community detection methods for multilayer networks \cite{roozbahani2021systematic} \cite{magnani2021community} \cite{huang2021survey} Despite these efforts, there is still a pressing need for a deeper exploration of the unique challenges posed by multilayer networks and for identifying emerging opportunities in this rapidly evolving field. This paper aims to fill that gap by introducing a taxonomy of existing methods, systematically analyzing the specific challenges in multilayer community detection, and proposing potential directions for future research. Additionally, we examine the practical applications of multilayer networks in real-world scenarios, demonstrating their significance across diverse domains.

The key contributions of this paper are as follows:
\begin{itemize}
    \item A comprehensive overview of community detection techniques tailored to multilayer networks.
    \item A detailed discussion of real-world applications of multilayer networks, showcasing their ability to model complex, multidimensional relationships.
    \item An in-depth analysis of current challenges in multilayer community detection, along with suggestions for innovative contributions to advance the field.
\end{itemize}

The structure of the paper is as follows: Section \ref{sec:problem description} presents an overview of the community detection problem in multilayer networks. Section \ref{sec: applications} discusses real-world applications of multilayer networks. Following that, we examine the various challenges, opportunities, and limitations associated with community detection in multilayer networks in Section \ref{sec: challenges}. Finally, Section \ref{sec:conclusion} wraps up the paper.

\section{\uppercase{community detection in Multilayer networks}}
\label{sec:problem description}

\begin{figure}[h!]
    \centering
    \includegraphics[width=\linewidth]{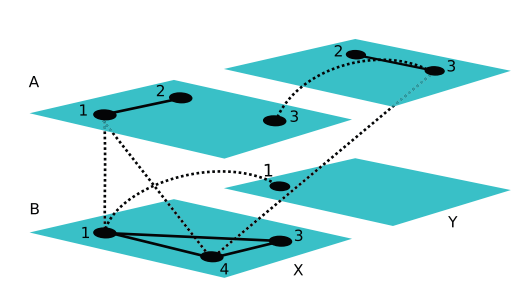}
    \caption{Example of multilayer networks.}
    \label{multilayer}
\end{figure}

Multi-layer networks (Figure \ref{multilayer}) are formed by combining or overlaying multiple individual networks that may be interconnected. In this framework, two fundamental definitions are commonly used to describe and analyze these complex structures. Kivela \textit{et al.} \cite{kivela2014multilayer} defines a multilayer network as a quadruplet $M=(V_M, E_M, V, L)$ where $V_M \subseteq V \times L_1 \times .... \times L_d$ is the set of node-layer combinations in which a node is present in the corresponding layer, and $E_M \subseteq V_M \times V_M$ is a set of edges connecting nodes in different layers. Conversely, Boccaletti \textit{et al.} \cite{boccaletti2014structure} defines a multilayer network as a couple $M = (G, C)$ where $G=\{G_\alpha; \alpha \in \{1, ..., d\}\}$ is a series of graphs $G_\alpha =(V_\alpha; E_\alpha)$ (called layers of $M$) and $C=\{E_{\alpha\beta} \subseteq V_\alpha \times V_\beta; \alpha, \beta \in \{1, ..., d\}, \alpha \neq \beta\}$ is the set of interconnections between nodes of different layers (called cross-layers).

Considering the design of multi-layer networks, various network types can be differentiated based on the nodes and links they include. This versatility allows for the exploration of diverse network structures and relationships, facilitating a more comprehensive analysis and understanding of complex systems. The main types of networks that can be distinguished include:
Here's a revised list with more consistent formatting and structure:

\begin{itemize}
    \item \textbf{Heterogeneous networks} \cite{sahneh2013effect}: Graphs in which each node represents a distinct type.
    \item \textbf{k-partite networks} \cite{kivela2014multilayer}: Networks with multiple disjoint node types, where connections occur only between nodes of different types, with no links within the same type.
    \item \textbf{Multi-relational networks} \cite{cai2005community}: Networks with a single node type but various link types, where each link represents a different kind of relationship between nodes.
    \item \textbf{Multislice networks} \cite{mucha2010community}: Networks that combine individual slices, each linked by connections that join corresponding nodes across slices.
    \item \textbf{Interdependent networks} \cite{sola2013eigenvector}: Composites of multiple layers, each layer representing a different type of relationship and possibly containing distinct sets of nodes \cite{wang2015evolutionary} \cite{gomez2012evolution} \cite{buldyrev2010catastrophic}.
    \item \textbf{Multiplex networks} \cite{boccaletti2014structure}: Sequences of graphs with a fixed set of nodes connected by various types of links.
\end{itemize}

\begin{figure}[h!]
    \centering
    \includegraphics[width=\linewidth]{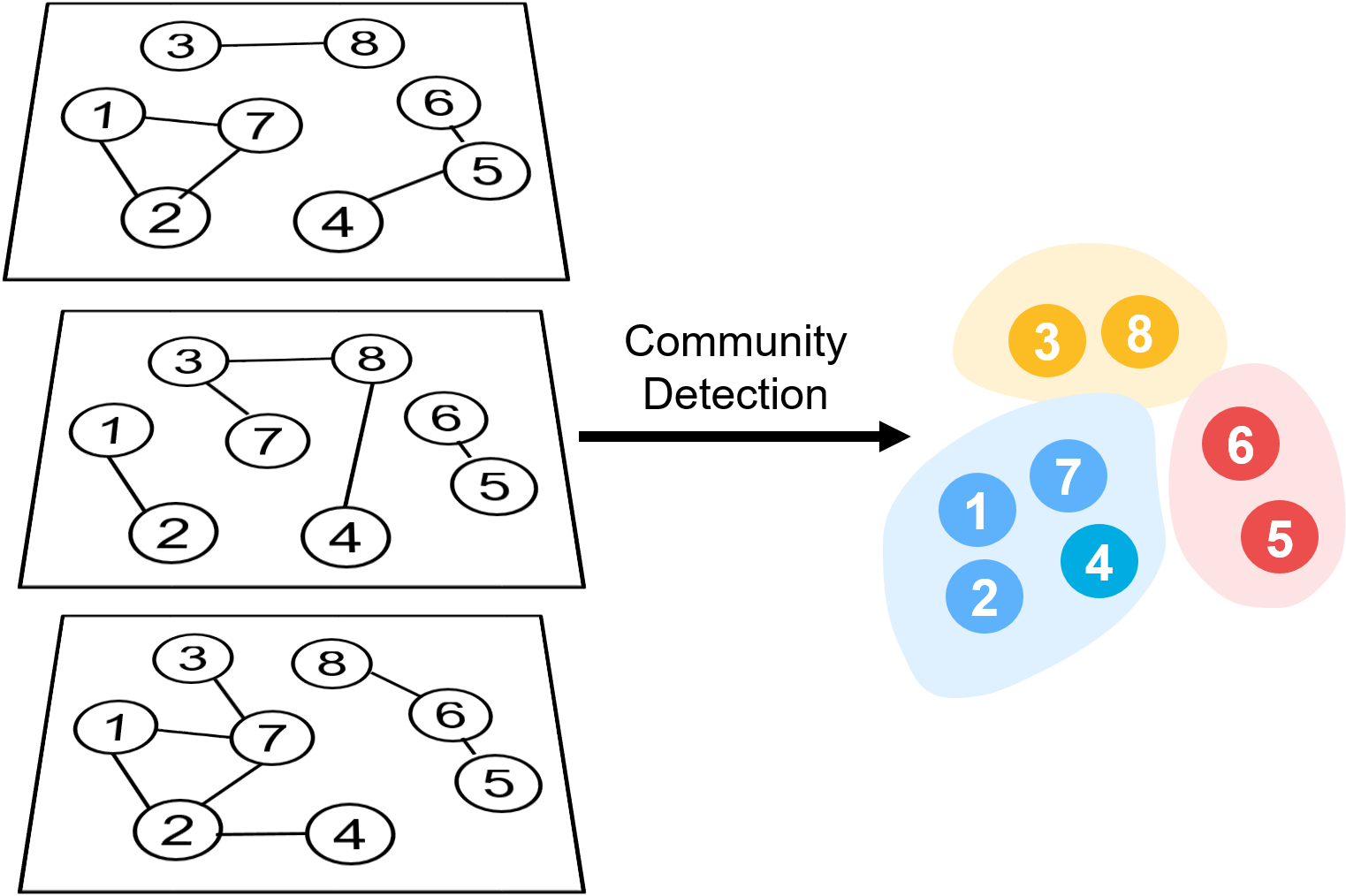}
    \caption{Example of community detection in multilayer networks.}
    \label{multiplex}
\end{figure}

Community structure plays a crucial role in social networks, significantly influencing our understanding of network dynamics. Within social networks, a community refers to sub-graphs where nodes exhibit stronger internal connections compared to their connections with the rest of the network \cite{girvan2002community}. Identifying such communities entails uncovering groups that share similar connectivity patterns within the network. Therefore, the problem of community detection requires the partitioning of the network into sub-groups or communities as illustrated in Fig.\ref{multiplex}. This partitioning is denoted as $C = \{C_1,...C_k\}$, where each element $C_{l}$ ($l = 1, 2,..., k$) is a proper subset of vertices $V$, and $k$ is the total number of communities. Initially, communities were treated as disjoint, with nodes assigned to separate communities \cite{girvan2002community}. In this framework, communities were distinct entities represented by fixed partitions. However, a new approach has emerged, allowing nodes to belong to multiple communities simultaneously, leading to the concept of overlapping communities \cite{palla2005uncovering}.

We can categorize existing approaches for community detection in multilayer networks into two hierarchical levels. The first level, data type, classifies methods into two broad categories based on the type of data used to define the community detection problem: structure-based and embedding-based approaches. The second level, resolution type, further differentiates techniques based on the conceptual framework they employ to detect communities. Notably, each data type supports multiple resolution types. The following section provides a detailed explanation of each category.

\subsection{Structure-based approaches}
Structure-based approaches to community detection focus on leveraging the network structure directly during the detection process. These methods do not rely on external or auxiliary information but instead use the relationships and patterns within the network itself. The structure of the network, which could be represented as a graph with nodes and edges, plays a central role in detecting communities. The structure-based approaches can be broken down into three main categories:

\begin{itemize}
    \item \textbf{Direct methods}:  
   These methods focus on analyzing the entire network as a single unified component to uncover community structures. Direct methods consider the global connectivity of the network and aim to identify communities by detecting patterns of connectivity (such as dense subgraphs or clusters) across the whole network. These approaches typically use algorithms that optimize a specific quality function, like modularity, to partition the network into communities. The methods try to extract a global community structure that simultaneously encompasses interactions from all layers.

   \item \textbf{Assembly methods}:  
   These methods take a more localized and layered approach, specifically designed for multilayer networks (networks with multiple layers of interactions). Instead of treating the entire network as a single unit, assembly methods detect communities separately within each individual layer of the multilayer network. Each layer may represent a different type of relationship (such as social interactions, collaborations, or communication channels). Once the community structure is identified in each layer, a consensus process is applied across all layers to find a common community structure. This step seeks to reconcile the different partitions obtained from each layer and combine them into a final partitioning of the entire multilayer network. 
    
   \item \textbf{Flattening methods}:  
   Flattening methods simplify the analysis by transforming the multilayer network into a single-layer network. Instead of working with multiple layers, this technique aggregates or collapses all layers into one unified network representation. The idea is to combine the information from all the layers, potentially by adding weighted edges or creating a composite graph that retains some of the interactions from the original layers. Once this single-layer network is obtained, traditional community detection algorithms can be applied \cite{boukabene2024flattening}. 
\end{itemize}

\subsection{Embedding-based approaches}

Embedding-based approaches transform the network into a latent space, where each node is represented as a low-dimensional vector. This transformation converts the community detection problem into a clustering problem, enabling the application of machine learning techniques to identify communities within the latent space. Such methods have gained prominence for their ability to handle the complexity of multilayer networks, as they effectively reduce the high-dimensional, interconnected structure into a simpler, computationally tractable form while preserving essential structural and relational properties.

Through a comprehensive review of the literature, we identified that existing algorithms for embedding generation in multilayer networks can be categorized into two main classes based on how they handle information from multiple layers during the embedding process: those that share information during the learning phase and those that do not. To better understand and organize these approaches, we propose a taxonomy that classifies embedding-based algorithms into two primary categories:

\begin{itemize}
    \item \textbf{Joint Representation Learning:}
   In this approach, network embedding techniques are applied independently to each layer of the multilayer network. After generating embeddings for individual layers, the representations are combined using fusion techniques to produce a unified embedding \cite{boukabene2025smart}. Fusion strategies may include averaging, concatenation, or more sophisticated methods that weigh layers differently based on their importance. 

   \item  \textbf{Coordinate Representation Learning:}
   Unlike joint representation learning, coordinate representation learning applies embedding techniques simultaneously across all layers, integrating information from each layer during the learning process. By jointly optimizing the embeddings, this approach creates a more comprehensive and cohesive representation of the multilayer network. It inherently captures the interplay between layers, making it well-suited for scenarios where inter-layer relationships are critical. 
\end{itemize}

\section{\uppercase{Real-world applications}}
\label{sec: applications}
In this section, we examine several real-world applications of multilayer networks, focusing specifically on transportation, finance, social networks, and biological networks, particularly within community detection.
\subsection{Transportation Networks}
Traffic dynamics represent a critical application of complex network theory, underscoring the importance of transportation systems in modern society. Traditionally, research has treated different modes of transportation, such as buses, subways, and airlines, as isolated systems, overlooking the complex interactions between them. However, recent studies have increasingly applied multilayer network models to capture these interconnections, offering deeper insights and optimization opportunities.

For instance, Du \textit{et al.} \cite{du2016physics} developed a two-layered traffic network model that uses particle swarm optimization to optimize passenger distribution, demonstrating that adjusting transfer costs can improve overall network capacity. Similarly, Hong and Liang \cite{hong2016analysis} applied a multilayer framework to analyze the Chinese airline network, revealing that it exhibits higher connectivity compared to other airlines. Ding \textit{et al.} \cite{ding2018detecting} studied Kuala Lumpur’s railway and urban street networks, showing that expanding rail infrastructure induces significant structural changes and enhances accessibility. In another study, Yildirimoglu and Kim\cite{yildirimoglu2018identification} integrated multiple transport modes into a three-layer urban traffic network, identifying geographically connected communities that provide valuable insights for urban planning and policy-making. These studies illustrate the power of multilayer network models in advancing our understanding and optimization of interconnected transportation systems.

\subsection{Finance Networks}

Community detection in multilayer networks plays a vital role in finance by uncovering clusters of interconnected entities, such as banks, investors, and assets, that exhibit similar behaviours or risks. By identifying these communities, analysts can evaluate systemic risk, understand contagion dynamics, and recognize market segmentation. This insight facilitates improved risk management, regulatory oversight, and more informed investment strategies. For instance, Biondo \textit{et al.} \cite{biondo2017informative} proposed a multilayer network model to simulate information diffusion and financial transactions. Similarly, Gao \textit{et al.} \cite{gao2024systemic} developed an approach to estimate China’s multilayer financial network based on various financial relationships among banks, assets, and firms, utilizing data from China’s banking system in 2021. In another study, Fernandes \textit{et al.} \cite{fernandes2019centrality} explored the relative importance of consumers by examining whether socioeconomic variables influence their centrality. They detected the communities within the network to which consumers belong, identified consumption patterns, and investigated the relationship between co-marketing and consumer choices.

\subsection{Social Network}

Community detection in multilayer networks has become a crucial focus in social network analysis, aimed at decoding the complex interactions among individuals to reveal structural patterns. With the expansion of online platforms like Facebook and Twitter, the demand to uncover community structures within these layered networks has grown significantly. Such analyses support applications like tracking information diffusion, understanding trade networks, and identifying key influencers, making multilayer approaches invaluable for capturing the dynamic nature of real-world interactions.

In international trade networks—often modelled as bipartite systems linking countries and industries—multilayer models have proven especially effective. For instance, Basaras \textit{et al.} \cite{basaras2017identifying} developed a technique for identifying influential spreaders. Following the rapid proliferation of social media and microblogging platforms in recent years, Oro \textit{et al.} \cite{oro2017detecting} introduced a novel approach for identifying users capable of influencing the choices of others. These studies underscore the need for advanced algorithms that can unlock insights from the vast and complex data generated by social networks, further enhancing our ability to analyze and interpret interconnected social and economic systems.

\subsection{Biological Networks}
Biological systems, from individual cells to the complexities of the human brain, are characterized by intricate interactions, making multilayer networks an effective representation of these phenomena \cite{guerrout2020hidden,guerrout2021image}. Central to many biological processes are densely connected networks of proteins, known as protein-protein interaction (PPI) networks, which elucidate how protein groups communicate and function together. 

Recent studies have highlighted the limitations of simple aggregated networks in capturing the full complexity of biological information. For instance, Chen \textit{et al.} \cite{chen2018parallel} developed the MLPCD algorithm, integrating Gene Expression Data with PPI networks to enhance the detection of protein complexes and functional modules through an improved Louvain algorithm\cite{blondel2008fast}. In network neuroscience, there is increasing emphasis on recognizing overlapping communities to better capture the complex organization of the brain. Zhang \textit{et al.} \cite{zhang2018central} introduced a central edge selection algorithm for community detection in PPI networks, demonstrating significant effectiveness across benchmark datasets. Additionally, research on human brain networks has progressed, with Cantini \textit{et al.} \cite{cantini2015detection} integrating various genomic data layers to identify cancer-related genes, and Sanchez-Rodriguez \textit{et al.} (2019) examining temporal scales within brain graphs to reveal modular organizational patterns. Recently, Comito \textit{et al.} \cite{comito2024integrating} proposed a multi-modal approach for the early diagnosis of Alzheimer’s disease (AD), which is crucial for ensuring timely treatment and care for patients. As advancements in network science continue to provide rich data resources, ongoing research is essential to further unravel the underlying patterns and functionalities of biological systems.

\section{\uppercase{Challenges and Future Directions}}
\label{sec: challenges}

Despite the significant advancements in the field of community detection in multilayer networks, there are still several gaps in the literature that warrant attention. In this section, we will address the various challenges and limitations associated with existing methods. Additionally, we will propose new research directions that can serve as a guide for future contributions to the field.

\subsection{Network Types}
While multiplex networks are the most widely researched type of multilayer networks, other important variants are frequently overlooked. These include networks that incorporate heterogeneous information \cite{cao2021knowledge}, as well as those with attributes on their nodes or edges \cite{wu2013multi}. Additionally, less common types such as signed \cite{xu2019link}, directed \cite{malliaros2013clustering}, or weighted networks \cite{nicolini2017community} receive limited attention. These diverse variants introduce added complexity and depth to network analysis, yet they necessitate advanced methodologies to thoroughly understand their unique properties and behaviors. This presents a valuable opportunity for new research, as exploring these lesser-studied network types could significantly enhance our understanding of multilayer networks and their applications.

\subsection{Network Dynamics}
Most research in multilayer networks tends to concentrate on static structures, often overlooking the dynamic aspects that provide a more accurate representation of real-world scenarios \cite{rossetti2018community,dakiche2019tracking}. This gap presents a promising area for growth, as understanding the dynamics of multilayer networks is crucial for capturing their true complexities. In this context, the work of Amelio \textit{et al.} \cite{amelio2017evolutionary} stands out, as they introduced one of the first approaches to address these dynamic elements. Their contributions mark an important step forward in the field and highlight the necessity for further exploration into how the dynamic nature of multilayer networks can influence community detection and other related analyses.

\subsection{Large-scale Networks}
Many current methods for community detection in multilayer networks are primarily tailored for small-scale networks, concentrating on enhancing detection quality while often overlooking the complexities presented by large-scale networks. These larger networks, which can include a vast number of nodes, edges, and communities, pose significant challenges for traditional detection techniques \cite{watts1998collective} \cite{barabasi2003scale}. To navigate these extensive networks, researchers frequently turn to dimension-reduction strategies, a promising yet underexplored area \cite{bhatia2018dfuzzy}. This gap in research underscores the need for innovative solutions that can effectively address the intricacies of large-scale multilayer networks, ultimately paving the way for more robust community detection methods that can operate at scale.

\subsection{Community Detection Process}
Community detection in multilayer networks involves three primary techniques: flattening, assembly, and direct approaches \cite{magnani2021community} \cite{huang2021survey} \cite{roozbahani2021systematic}. Flattening methods transform the multilayer network into a single-layer network and then apply traditional community detection algorithms. In contrast, assembly techniques identify communities within each layer and then combine them using a consensus method. Finally, direct approaches identify communities without any simplification of the network.

After reviewing the existing methods, we observed that direct approaches have received the most attention and analysis. There is a promising opportunity for future research in further exploring flattening and assembly techniques, as these areas are less studied. Utilizing similarity metrics \cite{wills2020metrics} to guide the resolution process may prove effective, as similarity metrics for multilayer networks is a nascent field that requires deeper investigation. Additionally, employing embedding techniques could enhance machine learning approaches for multilayer networks, such as the adaptive fusion mechanism proposed in \cite{wang2023graph}.

\subsection{Community Structure}
Overlapping communities in multilayer networks are still largely uncharted territory, providing a promising opportunity for new researchers. In these overlapping communities, nodes can belong to multiple groups  \cite{xie2013overlapping}, which reflects more realistic conditions and can yield deeper insights into the structure and dynamics of complex networks.

Community structures in multilayer networks can be divided into three main categories: local, global, and similarity-based communities. Most existing methods tend to concentrate on identifying global community structures \cite{mucha2010community}, frequently neglecting local \cite{berlingerio2011finding} and similarity-based communities \cite{brodka2013introduction}. Local communities, which emerge from node similarities and localized interactions, are essential for grasping the finer details of networks but remain underexplored in current research.

This unresolved issue highlights the need for comprehensive solutions and further research to develop methods that accommodate overlapping memberships while exploring both global and local community structures in multilayer networks. Tackling these challenges will greatly advance the field of community detection and improve our understanding of complex networks.

\subsection{Network Prior Information}
Community detection in real-life situations often faces the problem of dealing with mostly unlabeled data because collecting data can be expensive. This can make it hard to know how many communities exist within a network. Many algorithms, in particular clustering algorithms, require users to specify the number of communities beforehand, which limits their usefulness. Beldi \textit{et al.} \cite{beldi2023qlearning} proposed an approach that has great potential and could improve community detection methods in many areas. Finding ways to adapt these single-layer methods to automatically figure out the number of communities could help solve this issue.

\subsection{Multilayer Network Embeddings}
Most of the community detection techniques rely on analyzing network topology to identify clusters. However, the rise of node embedding techniques represents a significant advancement in this field. By transforming high-dimensional network data into lower-dimensional forms, these techniques maintain critical structural and relational information \cite{goyal2018graph}. This allows for a richer understanding of node connectivity, as they capture intricate relationships that enhance the clustering process \cite{wang2017computational}. 

Despite the promising field of node embeddings, there is a noticeable lack of techniques specifically tailored for multilayer networks, and the potential for hybrid approaches that integrate traditional methods with node embeddings has yet to be fully realized. This gap suggests an exciting direction for future research. Investigating how node embeddings can be effectively combined with conventional community detection methods in multilayer networks could lead to innovative techniques capable of addressing the unique complexities inherent in such structures. These developments could not only improve the accuracy of community detection but also provide deeper insights into the dynamics of multilayer networks, ultimately advancing our understanding of complex systems.

\section{\uppercase{Conclusion}}
\label{sec:conclusion}
Detecting communities in multilayer networks is a complex and challenging task, as it requires analyzing multiple layers of interconnected relationships between nodes. Despite these challenges, significant progress has been made, leading to the development of a variety of methods and algorithms aimed at uncovering community structures in such networks. This paper explores the different types of multilayer networks and their key applications in real-world scenarios. Additionally, it highlights the challenges faced in this field and proposes potential research directions to enhance community detection techniques and their effectiveness in multilayer contexts.

%
%
%
\bibliographystyle{ieeetr}

%

\end{document}